\documentclass{article}
\usepackage{amsfonts}
\usepackage{spconf,amsmath,graphicx,epsfig}
\usepackage{amsmath, amsthm, amssymb, epsfig}
\usepackage{xcolor}
\usepackage{hyperref}

\title{Exploiting Hybrid Models of Tensor-Train Networks for Spoken Command Recognition}
\name{Jun Qi$^{1}$,  Javier Tejedor$^{2}$ }
\address{$^1$Electrical and Computer Engineering, Georgia Institute of Technology, Atlanta, GA, USA \\
$^2$ Escuela Politecnica Superior, Universidad San Pablo-CEU, CEU Universities, Madrid, Spain}

\begin{document}
\ninept
\maketitle
\begin{abstract}
This work aims to design a low complexity spoken command recognition (SCR) system by considering different trade-offs between the number of model parameters and classification accuracy. More specifically, we exploit a deep hybrid architecture of a tensor-train (TT) network to build an end-to-end SRC pipeline. Our command recognition system, namely CNN+(TT-DNN), is composed of convolutional layers at the bottom for spectral feature extraction and TT layers at the top for command classification. Compared with a traditional end-to-end CNN baseline for SCR, our proposed CNN+(TT-DNN) model replaces fully connected (FC) layers with TT ones and it can substantially reduce the number of model parameters while maintaining the baseline performance of the CNN model. We initialize the CNN+(TT-DNN) model in a randomized manner or based on a well-trained CNN+DNN, and assess the CNN+(TT-DNN) models on the Google Speech Command Dataset. Our experimental results show that the proposed CNN+(TT-DNN) model attains a competitive accuracy of $96.31\%$ with $4$ times fewer model parameters than the CNN model. Furthermore, the CNN+(TT-DNN) model can obtain a 97.2$\%$ accuracy when the number of parameters is increased.
\end{abstract}
\begin{keywords}
Tensor-Train network, spoken command recognition, convolutional neural network
\end{keywords}
\section{Introduction}
The state-of-the-art deep learning systems like spoken command recognition (SCR) highly rely on the use of a great amount of training data and computing power from centralized cloud services~\cite{yu2016automatic}. This is attributed to the fact that over-parameterized deep neural networks can simply find satisfied local optimal points which results in better performance~\cite{neyshabur2019towards}. On the other hand, a practical low-complexity SCR system could bring about many benefits and make new applications, like mobile-based speech assistants~\cite{matarneh2017speech} run on users' phones without sending requests to a remote server when they need access to a deep learning model. In doing so, a localized low-complexity SCR system on local devices can possibly maintain the classification accuracy. Therefore, it motivates us to work on new algorithms and architectures to design a low-complexity SCR system by shrinking existing deep learning models without losing their capabilities in terms of representation and generalization powers.

The approaches to shrinking deep learning models can be divided into two categories. One refers to the use of new deep learning architectures, such as convolutional neural network (CNN) with various novel structures~\cite{he2016deep}; the second class is related to model pruning or sparseness methodologies~\cite{liu2018rethinking, zhu2017prune}, where either theoretically guaranteed pruning method or randomized lottery ticket hypothesis~\cite{malach2020proving, frankle2019stabilizing} can be used to attain a tiny machine learning system. Our work focuses on the first class by proposing a hybrid architecture based on a tensor-train (TT) network.

TT is related to a tensor decomposition and is capable of characterizing the representation of a chain-like product of three-index core tensors for a high-order tensor~\cite{oseledets2011tensor}. In doing so, the overhead of memory storage can be greatly reduced. Compared with other tensor decomposition techniques~\cite{sidiropoulos2017tensor} like Tucker decomposition (TD)~\cite{kim2007nonnegative} and CANDECOMP/PARAFAC decomposition~\cite{phan2011parafac, faber2003recent}, TT is a special case of a tree-structured tensor network and can be simply scaled to arbitrarily high-order tensors~\cite{novikov2015tensorizing}. Thus, TT has been widely applied in signal processing and machine learning domains, such as video classification~\cite{yang2017tensor} and deep learning recommendation systems~\cite{deng2019tie}. In particular, our previous work on speech enhancement~\cite{qi2020tensor, Qi2020} shows that the TT representation for deep neural network (DNN), namely TT-DNN, is employed to maintain the DNN baseline results, but it owns substantially much fewer model parameters.

In this work, we further investigate the use of TT to compose a hybrid end-to-end architecture combining convolutional layers with TT layers, where the convolutional layers are placed at the bottom, and the TT layers are assigned at the top. The convolutional layers are used to transform time-series speech signals into spectral features, and the TT layers are responsible for classification. A TT-DNN with multiple TT layers can significantly reduce the number of model parameters, while it maintains the baseline performance. More importantly, our previous work~\cite{qi2019theory, qi2020mean, qi2020analyzing} shows that the better representation power of DNN requires greater width of fully connected (FC) layers, which implies huge computational resources and storage cost. However, a localized device cannot generally provide a powerful hardware platform to support an even over-parameterized neural network.

Furthermore, although our previous work suggests that the singular value decomposition cannot keep the baseline performance, such an over-parameterized network can be redesigned into a TT representation and the related model size can be further reduced into a small scale. Thus, in this work, we investigate the use of TT to compose a hybrid architecture, namely CNN+(TT-DNN), for SCR. Moreover, we assess the performance of CNN+(TT-DNN) on the task of spoken command recognition, which is a limited-vocabulary but reasonably challenging speech recognition task.

In this work, we aim to provide the following contributions:
\begin{enumerate}
\item We propose a hybrid end-to-end SCR architecture with convolutional layers and TT layers.
\item Different model initialization strategies are compared.
\item We compare our TT solution with the more classical Tucker decomposition on FC layers.
\end{enumerate}


\section{Tensor-Train Network}


\subsection{Tensor-Train Decomposition}
\begin{figure}
\centerline{\epsfig{figure=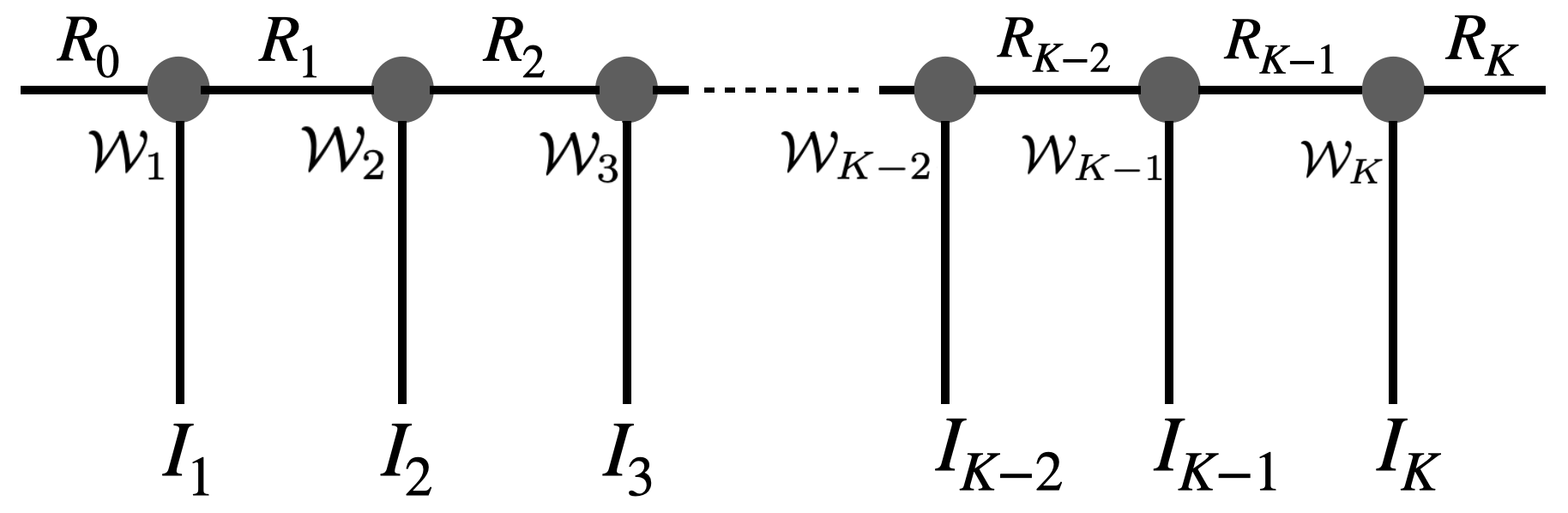, width=65mm}}
\caption{{\it An illustration of TTD, which is a tensor of order $K$ in TT format, where the core tensors are of order $3$. Given a set of TT ranks $\{R_{0}, R_{1}, ..., R_{K}\}$, a circle represents a core tensor $\mathcal{W}_{k} \in \mathbb{R}^{I_{k} \times R_{k-1} \times R_{k}}$, and each line is associated with the dimension.}}
\label{fig:ttd}
\end{figure}

We first define some useful notations: $\mathbb{R}^{d}$ denotes a $d$-dimensional real space, $\mathbb{R}^{I_{1} \times I_{2} \times \cdot\cdot\cdot I_{K}}$ is a space of $K$-order tensors, $\mathcal{W} \in \mathbb{R}^{I_{1} \times \cdot\cdot\cdot \times I_{K}}$ is a $K$-dimensional tensor, and $\textbf{W} \in \mathbb{R}^{I_{1} \times I_{2}}$ is an $I_{1} \times I_{2}$ matrix. Additionally, the symbol $[K]$ means a set of integers $\{1, 2, ..., K\}$. 

As shown in Figure~\ref{fig:ttd}, the technique of tensor-train decomposition (TTD) assumes that given a set of TT-ranks $\{R_{0}, R_{1}, ..., R_{K}\}$, a $K$-order tensor $\mathcal{W}\in \mathbb{R}^{I_{1}\times I_{2}\times \cdot\cdot\cdot\times I_{K}}$ is factorized into the multiplication of 3-order tensors $\mathcal{W}_{k} \in \mathbb{R}^{I_{k} \times R_{k-1} \times R_{k}}$. More specifically, given a set of indices $\{i_{1}, i_{2}, ..., i_{K}\}$, $\mathcal{W}(i_{1}, i_{2}, ..., i_{K})$ is decomposed as Eq.~(\ref{eq:tt}). 
\begin{equation}
\label{eq:tt}
\mathcal{W}(i_{1}, i_{2}, ..., i_{K}) = \prod\limits_{k=1}^{K} \mathcal{W}_{k}(i_{k}),
\end{equation}
where $\forall i_{k} \in I_{k}$, $\mathcal{W}_{k} \in \mathbb{R}^{I_{k} \times R_{k-1}\times R_{k}}$ and $\mathcal{W}_{k}(i_{k}) \in \mathbb{R}^{R_{k-1} \times R_{k}}$. Since $R_{0} = R_{K} = 1$, the term $\prod_{k=1}^{K} \mathcal{W}_{k}(i_{k})$ is a scalar value.

One example of TTD is shown as follows: for a $3$-order tensor $\mathcal{W}(i_{1}, i_{2}, i_{3}) = i_{1} + i_{2} + i_{3}$, given the set of TT ranks $\{1, 2, 2, 1\}$, the use of TTD on $\mathcal{W}$ outputs $3$ core tensors as Eq.~(\ref{eq:2}). 
\begin{equation}
\label{eq:2}
\mathcal{W}_{1}[i_{1}] := [ \begin{matrix} i_{1} & 1  \end{matrix}],  \mathcal{W}_{2}[i_{2}] := \left[\begin{matrix} 1 & 0 \\ i_{2} & 1 \end{matrix}\right], \mathcal{W}_{3}[i_{3}] := \left[ \begin{matrix} 1 \\ i_{3} \end{matrix} \right], 
\end{equation}
which is derived from Eq.~(\ref{eq:ww}). 
\begin{equation}
\label{eq:ww}
\mathcal{W}(i_{1}, i_{2}, i_{3}) = [\begin{matrix} i_{1} & 1  \end{matrix}] \left[\begin{matrix} 1 & 0 \\ i_{2} & 1 \end{matrix}\right] \left[ \begin{matrix} 1 \\ i_{3} \end{matrix} \right] = i_{1} + i_{2} + i_{3}.
\end{equation}


\subsection{Tensor-Train Deep Neural Network}
\begin{figure}
\centerline{\epsfig{figure=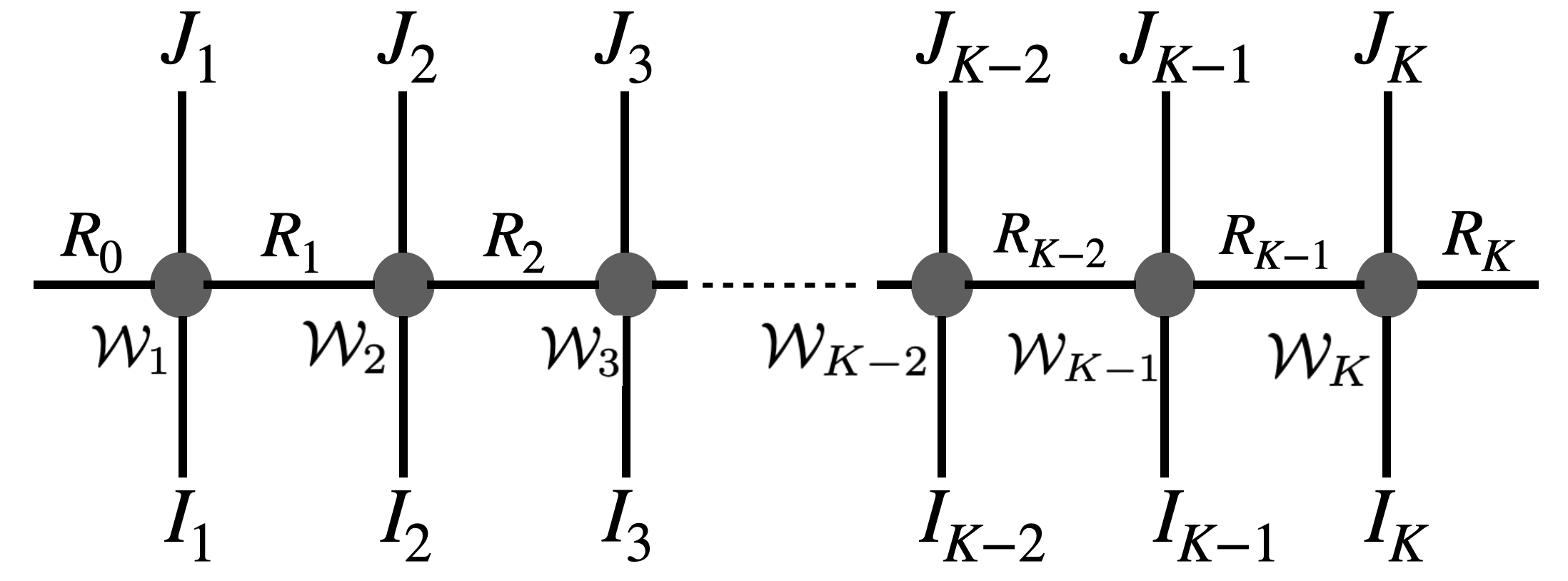, width=65mm}}
\caption{{\it An illustration of TTN, which is a tensor of order $K$ in TT format, where the core tensors are of order $4$. Given a set of TT ranks $\{R_{0}, R_{1}, ..., R_{K}\}$, a circle represents a core tensor $\mathcal{W}_{k} \in \mathbb{R}^{I_{k} \times J_{k} \times R_{k-1} \times R_{k}}$, and each line is associated with the dimension.}}
\label{fig:ttnn}
\end{figure}

A tensor-train network (TTN) refers to a TT representation of a feed-forward neural network with a FC hidden layer. In more detail, for an input tensor $\mathcal{X} \in \mathbb{R}^{I_{1} \times I_{2} \times \cdot\cdot\cdot \times I_{K}}$ and an output tensor $\mathcal{Y} \in \mathbb{R}^{J_{1} \times J_{2} \times \cdot\cdot\cdot \times J_{K}}$, we get Eq.~(\ref{eq:yy}) as follows:
\begin{equation}
\label{eq:yy}
\begin{split}
&\hspace{4mm} \mathcal{Y}(j_{1}, j_{2}, ..., j_{K})  \\
&= \sum\limits_{i_{1}=1}^{I_{1}} \cdot\cdot\cdot \sum\limits_{i_{K} = 1}^{I_{K}} \mathcal{W}((i_{1}, j_{1}), ..., (i_{K}, j_{K})) \cdot \mathcal{X}(i_{1}, i_{2}, ..., i_{K}) \\
&= \sum\limits_{i_{1}=1}^{I_{1}} \cdot\cdot\cdot \sum\limits_{i_{K} = 1}^{I_{K}} \left( \prod\limits_{k=1}^{K} \mathcal{W}_{k}(i_{k}, j_{k}) \right) \cdot \mathcal{X}(i_{1}, i_{2}, ..., i_{K}) \\
&= \sum\limits_{i_{1}=1}^{I_{1}} \cdot\cdot\cdot \sum\limits_{i_{K} = 1}^{I_{K}} \left( \prod\limits_{k=1}^{K} \mathcal{W}_{k}(i_{k}, j_{k}) \right) \cdot \prod\limits_{k=1}^{K} \mathcal{X}_{k}(i_{k})		 \\
&= \prod\limits_{k=1}^{K} \left( \sum\limits_{i_{k}=1}^{I_{k}} \mathcal{W}_{k}(i_{k}, j_{k}) \cdot \mathcal{X}_{k}(i_{k})  \right)		\\
&= \prod\limits_{k=1}^{K} \mathcal{Y}_{k}(j_{k}),
\end{split}
\end{equation}
where $\mathcal{X}_{k}(i_{k}) \in \mathbb{R}^{R_{k-1} \times R_{k}}$, and $\mathcal{Y}_{k}(j_{k}) \in \mathbb{R}^{R_{k-1} \times R_{k}}$ which results in a scalar $\prod_{k=1}^{K} \mathcal{Y}_{k}(j_{k})$ because of the ranks $R_{0} = R_{K} = 1$; $\mathcal{W}((i_{1}, j_{1}), (i_{2}, j_{2}), ..., (i_{K}, j_{K}))$ is closely associated with $\mathcal{W}(m_{1}, m_{2}, ..., m_{K})$ as defined in Eq.~(\ref{eq:tt}), if the indexes $m_{k} = i_{k} \times j_{k}, k\in [K]$ are set. As shown in Figure~\ref{fig:ttnn}, given the TT-ranks $\{R_{0}, R_{1}, ..., R_{K}\}$, the multi-dimensional tensor $\mathcal{W}$ is decomposed into the multiplication of $4$-order tensors $\mathcal{W}_{k} \in \mathbb{R}^{J_{k} \times I_{k} \times R_{k-1} \times R_{k}}$. Besides, to configure a TTN, the ReLU activation is applied to $\mathcal{Y}(j_{1}, j_{2}, ..., j_{K})$ as Eq.~(\ref{eq:6}).
\begin{equation}
\label{eq:6}
\begin{split}
\mathcal{\hat{Y}}(j_{1}, j_{2}, ..., j_{K}) &= \text{ReLU}(\mathcal{Y}(j_{1}, j_{2}, ..., j_{K})) \\
&= \text{ReLU}\left(\prod\limits_{k=1}^{K} \mathcal{Y}_{k}(j_{k})\right).
\end{split}
\end{equation}



\begin{figure}
\centerline{\epsfig{figure=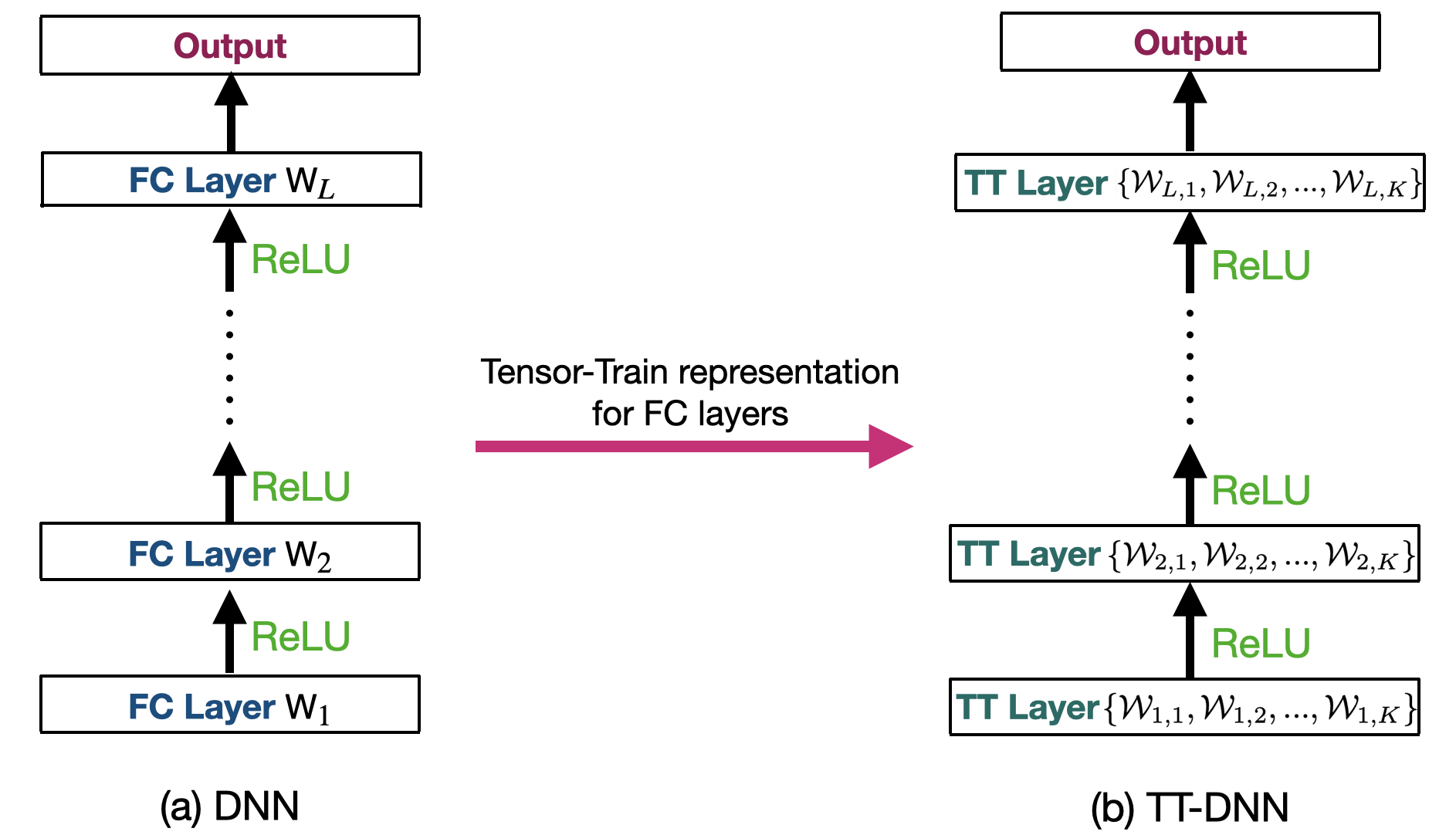, width=80mm}}
\caption{{\it Converting DNN into TT-DNN: setting up and training from scratch, or directly converting from a well-trained DNN. Each DNN weight $\textbf{W}_{l}$ is connected to $K$ core tensors $\mathcal{W}_{l, 1}, \mathcal{W}_{l, 2}, ..., \mathcal{W}_{l, K}$. }}
\label{fig:ttn}
\end{figure}

Then, TTN can be generalized to a deeper architecture and is closely associated with the TT representation for DNN, namely TT-DNN. Figure~\ref{fig:ttn} illustrates that a DNN model can be transformed into a TT-DNN structure, where all FC hidden layers can be represented as the corresponding TT layers. An additional softmax operation is imposed on the top layer for classification, but the operation is set as linear for regression problems. More specifically, $\forall k\in [K]$ and $\forall l\in [L]$, the DNN matrix $\textbf{W}_{l} \in \mathbb{R}^{J_{l} \times I_{l}}$ can be decomposed into $K$ core tensors $\{\mathcal{W}_{l, 1}, \mathcal{W}_{l, 2}, ..., \mathcal{W}_{l, K}\}$, where $\mathcal{W}_{l, k} \in \mathbb{R}^{J_{l,k} \times I_{l,k} \times R_{k-1} \times R_{k}}$, $J_{l}=J_{l, 1} \times J_{l, 2} \times \cdot\cdot\cdot \times J_{l, K}$ and $I_{l} = I_{l, 1} \times I_{l, 2} \times \cdot\cdot\cdot \times I_{l, K}$.

The TTD admits a TT-DNN with much fewer number of model parameters than the related DNN. More specifically, a DNN with $\sum_{l=1}^{L}J_{l} I_{l}$ parameters can be converted to a TT-DNN with fewer model parameters such as $\sum_{l=1}^{L}\sum_{k=1}^{K} J_{l, k}I_{l, k} R_{k-1}R_{k}$.

\subsection{Training TTN and TT-DNN Models}

In this work, we take into account two kinds of training methods for attaining a well-trained TTN or a TT-DNN. The first one randomly initializes the TT parameters and then trains the model parameters from scratch; the second one generates the TT models by applying TTD in well-trained CNN+DNNs followed by further fine-tuning of the model parameters. The first training method is preferred because the TT models can be independent of the related DNNs, but we also attempt to observe the model performance by using the second training methodology.

\section{CNN+(TT-DNN) for Spoken Command Recognition}
\begin{figure}[htp]
\centerline{\epsfig{figure=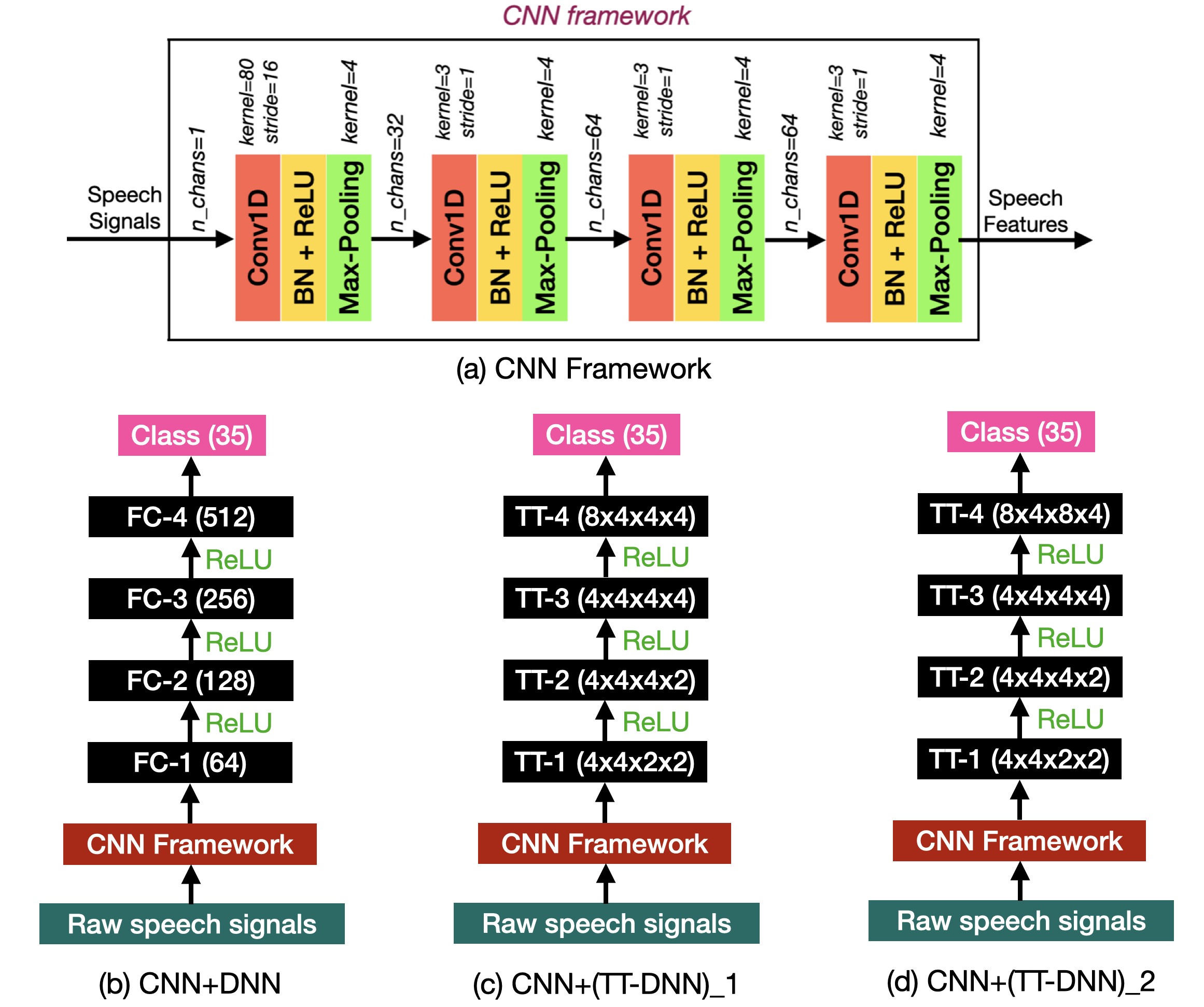, width=75mm}}
\caption{{\it CNN+DNN and CNN+(TT-DNN) for spoken command recognition, where the tensor shape $4\times 4 \times 2 \times 2$ refers to the $4$-order tensor in the space $\mathbb{R}^{4\times 4\times 2\times 2}$. CNN+(TT-DNN)$\_1$ and CNN+(TT-DNN)$\_2$ differ in the tensor shape of the top hidden layer. }}
\label{fig:exp}
\end{figure}

Figure~\ref{fig:exp} illustrates the proposed SCR system. The CNN framework is used to convert speech signals into spectral features in Figure~\ref{fig:exp} (a). The entire CNN framework consists of $4$ components, each of which is constructed by stacking $1$D convolutional layer with batch normalization (BN) and the ReLU activation, which is followed by a max-pooling layer with a kernel size of $4$. Particularly, the first component owns a kernel size of $80$ and $16$ strides, while kernel sizes of $3$ and $1$ stride are assigned to the other components. Moreover, the number of channels for the CNN framework follows the pipeline of $1$-$32$-$64$-$64$.

The spectral features associated with the outputs of the CNN framework are fed into the FC layers or TT layers, which are shown in Figure~\ref{fig:exp} (b) and (c), respectively. We set our baseline system as the CNN+DNN architecture in which several FC layers are stacked on top of the CNN layers. On the other hand, the FC layers are changed to the TT layers in our proposed CNN+(TT-DNN) model. Besides, two training methods for the CNN+(TT-DNN) model are considered: one refers to setting up the CNN+(TT-DNN) model with random initialization for model parameters, and another is derived from the TT decomposition of a well-trained CNN+DNN model. Moreover, the output of the SCR systems is connected to classification labels.

\section{Experiments}

\subsection{Data profile}
Our spoken command recognition experiments were conducted on the Google Speech Command dataset~\cite{warden2018speech}, which includes $35$ spoken commands, e.g., [`left', `go', `yes', `down', `up', `on', `right', ...]. There are a total of $11,165$ development and $6,500$ test utterances. The development data are randomly split into two parts: $90\%$ is used for model training and $10\%$ is used for validation. All the audio files are about $1$ second long, downsampled from $16$KHz to $8$KHz. The batch size was set to $256$ in the training process, and the speech signals in a batch were configured as the same length by zero-padding.

\subsection{Experimental setup}

The model architectures for the tested SCR systems are shown in Figure~\ref{fig:exp}, where we take three acoustic models into account. Figure~\ref{fig:exp} (b) shows our baseline SCR system in which $4$ FC layers ($64$-$128$-$256$-$512$) are stacked and $35$ classes are appended as the label layer. Figure~\ref{fig:exp} (c) illustrates our CNN+(TT-DNN) model where $4$ TT layers follow the tensor shape of $4\times 4\times 2\times 2 - 4\times 4\times 4\times 2 - 4\times 4\times 4\times 4 - 8\times 4\times 4\times 4$, whereas in Figure~\ref{fig:exp} (d) the shape of the top TT layer is modified to $8\times 4\times 8\times 4$ because a larger CNN+(TT-DNN) model is considered to compare SCR performances.

The loss objective function was based on the criterion of cross-entropy (CE), and the Adam optimizer with a learning rate of $0.01$ was used in the training stage. The loss value of CE is a direct assessment for the model performance of SCR, and Accuracy (Acc.) is an indirect measurement to evaluate speech recognition performance. There are a total of 100 epochs used in the training process. We report the best average accuracy for each SCR architecture with 10 runs. 

\subsection{Experimental results}

Our experiments are divided into three parts: (1) we set up the CNN+(TT-DNN) model with randomly initialized model parameters; (2) our proposed CNN+(TT-DNN) models are derived from a well-trained CNN+DNN; (3) we compare the use of TTD to the DNN model with Tucker decomposition in the DNN model.

\subsubsection{Randomly Initialized CNN+(TT-DNN)}

The CNN+(TT-DNN) models are randomly initialized, and CNN+(TT-DNN)$\_$1 and CNN+(TT-DNN)$\_$2 correspond to the models in Figure~\ref{fig:exp} (c) and (d), respectively.
Our models are compared with neural network models available in the literature, namely: DenseNet-121 benchmark used in~\cite{mcmahan2018listening} for SCR, Attention-RNN~\cite{de2018neural, yang2020characterizing}, which refers to the neural attention model, and QCNN~\cite{yang2021decentralizing}, which refers to the use of quantum convolutional features for the task. We extend the 10 classes training setup used in~\cite{yang2021decentralizing} to 35 classes to report its final results. All deployed models are trained with the same SCR dataset from scratch without any data augmentation~\cite{berg2021keyword} or pre-training techniques~\cite{seo2021wav2kws} to make a fair architecture-wise study.

\begin{table}[tpbh]\footnotesize
\center
\renewcommand{\arraystretch}{1.3}
\begin{tabular}{|c||c|c|c|c|}
\hline
Models      		& Params	$(\text{Mb})$		&   CE     	& Acc.  ($\%$)   \\
\hline
DenseNet-121~\cite{mcmahan2018listening}				&   7.978		&   0.473 	 	&    82.11			\\
\hline
Attention-RNN~\cite{de2018neural}				&   0.170		&   0.291 	 	&    93.90 			\\
\hline
QCNN~\cite{yang2021decentralizing}				&   0.186		&   0.280 	 	&    94.23			\\
\hline
CNN+DNN			&   0.216		&   0.251 	 	&    94.42 			\\
\hline
CNN+(TT-DNN)$\_$1	&   \textbf{0.056 }		&   0.137		&   96.31			\\
\hline
CNN+(TT-DNN)$\_$2	& 0.083   &  \textbf{0.124}	&  \textbf{97.20}	\\
\hline
\end{tabular}
\caption{The experimental results on the test dataset. Params. represents the number of model parameters; CE means the cross-entropy; and Acc. refers to the classification accuracy.}
\label{tab:tab3}
\end{table}

Experimental results are summarized in Table~\ref{tab:tab3}. The CNN+DNN model is taken as the baseline SCR system, and it attains $94.42\%$ accuracy and $0.251$ CE score, which are better than the results of the DenseNet model, the neural attention model, and the QCNN model. The CNN+(TT-DNN)$\_$1 model obtains better results in terms of accuracy ($96.31\%$ vs. $94.42\%$), CE (0.137 vs. 0.251), and it owns a much smaller model size (0.056 vs. 0.216) than the CNN+DNN model. More significantly, the CNN+(TT-DNN)$\_$1 model also outperforms the DenseNet, neural attention, and QCNN models in terms of smaller model size, lower CE value, and higher classification accuracy. The CNN+(TT-DNN)$\_$2 model, which owns more model parameters than the CNN+(TT-DNN)$\_$1 model ($0.083$ vs. $0.056$), obtains higher accuracy ($97.20\%$ vs. $96.31\%$) and lower CE value ($0.124$ vs. $0.137$) than the CNN+(TT-DNN)$\_1$ model. In other words, the CNN+(TT-DNN)$\_2$ model boosts a $9.2\%$ relative performance gain, and the CNN+(TT-DNN)$\_$1 model outperforms the rest of the systems (except CNN+(TT-DNN)$\_2$) for all metrics. 

\subsubsection{CNN+(TT-DNN) Based on A Well-Trained DNN}

Instead of using randomly initialized neural network parameters, we initialize our CNN+(TT-DNN) model from a well-trained CNN+DNN. Based on the CNN+DNN model in Table~\ref{tab:tab3}, the well-trained FC layers are transformed into the TT format, and the back-propagation algorithm is applied to further fine-tune the model parameters. We denote CNN+(TT-DNN)$\_$11 as the randomly initialized CNN+(TT-DNN) model, and CNN+(TT-DNN)$\_$12 refers to the TTD to the FC layers of the well-trained CNN+DNN model. In particular, the tensor shape of the CNN+(TT-DNN)$\_11$ and CNN+(TT-DNN)$\_12$ models is identical to that of the CNN+(TT-DNN)$\_1$ model as shown in Figure~\ref{fig:exp} (c).

\begin{table}[tpbh]\footnotesize
\center
\renewcommand{\arraystretch}{1.3}
\begin{tabular}{|c||c|c|c|c|}
\hline
Models      & Params	$(\text{Mb})$	&   CE     	&    Acc.  ($\%$)     			 \\
\hline
CNN+(TT-DNN)$\_$11	&   0.056 		&   0.137		&   96.31			 \\
\hline
CNN+(TT-DNN)$\_$12	&   0.056		&  \textbf{0.131}		&   \textbf{96.42}			 \\
\hline
\end{tabular}
\caption{The experimental results on the test dataset. CNN+(TT-DNN)$\_$11 is randomly initialized; CNN+(TT-DNN)$\_$12 denotes a CNN+(TT-DNN) derived from a well-trained CNN+DNN.}
\label{tab:tab4}
\end{table}

Table~\ref{tab:tab4} compares the performance of CNN+(TT-DNN)$\_$11 and CNN+(TT-DNN)$\_$12 models. Although the two models are set with the same number of parameters, CNN+(TT-DNN)$\_$12 can sightly improve the performance of CNN+(TT-DNN)$\_$11 by obtaining higher accuracy ($96.42\%$ vs. $96.31\%$) and lower CE ($0.131$ vs. $0.137$). The experiments suggest that a CNN+(TT-DNN) model derived from a well-trained CNN+DNN can outperform the randomly initialized CNN+(TT-DNN) model. 

\subsubsection{Comparison with Tucker Decomposition on DNN}

Tucker decomposition is a high-order extension to the singular value decomposition by computing the orthonormal spaces associated with the different modes of a tensor. Moreover, the CANDECOMP/PARAFAC decomposition can be taken as a special case of Tucker decomposition. Therefore, it is meaningful to verify whether Tucker decomposition applied to each FC layer can result in a model parameter reduction with only a small drop in performance.

We apply Tucker decomposition to the well-trained baseline model based on CNN+DNN and denote the related model as CNN+(TD-DNN). Table~\ref{tab:tab5} shows the results of CNN+(TT-DNN)$\_$2 and CNN+(TD-DNN) models, along with those of the baseline model (i.e., CNN+DNN). Although the CNN+(TD-DNN) model can slightly decrease the number of model parameters of the CNN+DNN, the SCR performance can be significantly degraded in terms of higher CE (0.286 vs. 0.251) and much lower accuracy ($91.25\%$ vs. $94.42\%$). Moreover, the results of the CNN+(TD-DNN) model are consistently worse than those of the CNN+(TT-DNN)$\_2$ model. Our results suggest that Tucker decomposition on DNN cannot maintain the baseline performance as the TT technique.

\begin{table}[tpbh]\footnotesize
\center
\renewcommand{\arraystretch}{1.3}
\begin{tabular}{|c||c|c|c|c|}
\hline
Models      & Params	$(\text{Mb})$	&   CE     	&    Acc.  ($\%$)     			 \\
\hline
CNN+DNN			&   0.216		&   0.251 	 	&    94.42 			\\
\hline
CNN+(TT-DNN)$\_$2	& 0.083             &   0.124		& 97.20	\\			
\hline
CNN+(TD-DNN)		&   0.206		&  0.286		&   91.25			 \\
\hline
\end{tabular}
\caption{A comparison of CNN+(TT-DNN)$\_$2 with CNN+(TD-DNN) and the baseline CNN+DNN models on the test dataset. }
\label{tab:tab5}
\end{table}

\vspace{-2mm}
\subsection{Summary}
Our experiments assess different model settings on the task of spoken command recognition. Our experimental results show that the TTD can maintain and even obtain better results than the given baseline models (DenseNet, neural attention, and QCNN models), when the CNN+(TT-DNN) model is initialized randomly or derived from a well-trained CNN+DNN. In particular, the performance of the CNN+(TT-DNN) model can be boost by increasing the number of the model parameters. Besides, Tucker decomposition cannot maintain the DNN baseline results as CNN+(TT-DNN) models.

\section{Conclusions}

This work investigates the implementation of a low-complexity SCR system by applying the TT technique. A hybrid model, namely CNN+(TT-DNN), is proposed to realize an end-to-end SCR pipeline, and the model can be randomly initialized or derived from a well-trained CNN+DNN. Our experiments were conducted on the Google Speech Command dataset for the task of spoken command recognition. The experimental results suggest that the CNN+(TT-DNN) model is able to obtain better performance than the CNN+DNN model, whereas Tucker decomposition on the CNN+DNN model reduces the performance. In particular, our proposed CNN+(TT-DNN) model with more parameters can further boost performance. The proposed TT technique could also benefit to large-scale architecture and pre-training methods for future studies. 

\clearpage
\bibliographystyle{IEEEbib}
\bibliography{refs}

\end{document}